\documentclass[12pt]{article}
\usepackage[left]{lineno}
\usepackage{blindtext}
\usepackage{enumitem}
\usepackage{setspace}
\usepackage{natbib}
\usepackage{graphicx}
\usepackage{epstopdf}
\usepackage[margin=2cm]{geometry}
\usepackage{hyperref}
\usepackage{setspace}
\usepackage{xcolor}
\usepackage{subfig}
\usepackage[us,12hr]{datetime}
\usepackage[version=3]{mhchem}
\usepackage{hyperref}
\usepackage{caption}
\usepackage{array}
\newcolumntype{P}[1]{>{\centering\arraybackslash}p{#1}}
\newcommand{\footremember}[2]{%
   \footnote{#2}
    \newcounter{#1}
    \setcounter{#1}{\value{footnote}}%
}
\newcommand{\footrecall}[1]{%
    \footnotemark[\value{#1}]%
}

\setcitestyle{super,open={},close={}}

\usepackage{ulem}
\usepackage{xcolor}

\begin{document}

\title{\Large{\bf{A Nineteen Day Earth Tide Measurement with a MEMS Gravimeter}}}
\author{
    \footnotesize{A Prasad}\footremember{Physics}{S.U.P.A., University of Glasgow, School of Physics and Astronomy, Kelvin Building, University Avenue, Glasgow, G12 8QQ, U.K.} 
\and \small{\hspace{-0.6cm}R. P. Middlemiss,}\footremember{Engineering}{University of Glasgow, James Watt School of Engineering, Rankine Building, Oakfield Avenue, Glasgow, G12 8LT, U.K.}
\and \small{\hspace{-0.6cm}K. Anastasiou,}\footrecall{Physics}
\and \small{\hspace{-0.6cm}S. G. Bramsiepe,}\footrecall{Physics}
\and \small{\hspace{-0.6cm}A. Noack,}\footrecall{Physics}
\and \small{\hspace{-0.6cm}D. J. Paul,}\footrecall{Engineering}
\and \small{\hspace{-0.6cm}K. Toland,}\footrecall{Physics}
\and \small{\hspace{-0.6cm}P. R. Utting,}\footrecall{Engineering}\hspace{0.1cm}\footrecall{Physics}
\and \small{\hspace{-0.6cm}and G. D. Hammond}\footrecall{Physics}}

\date{} 

\maketitle

\section*{Abstract}
The measurement of tiny variations in local gravity enables the observation of subterranean features. Gravimeters have historically been extremely expensive instruments, but usable gravity measurements have recently been conducted using MEMS (microelectromechanical systems) sensors. Such sensors are cheap to produce, since they rely on the same fabrication techniques used to produce mobile phone accelerometers. A significant challenge in the development of MEMS gravimeters is maintaining stability over long time periods, which is essential for long term monitoring applications. A standard way to demonstrate gravimeter stability and sensitivity is to measure the periodic elastic distortion of the Earth due to tidal forces - the Earth tides. Here we present a nineteen day measurement of the Earth tides, with a correlation coefficient to the theoretical signal of 0.979. The estimated bias instability of the proposed gravimeter is 8.18 $\mu$Gal for an averaging time of $\sim$400 s when considering the raw, uncompensated data. The bias instability extracted from the sensor electronic noise sits just under 2 $\mu$Gal for an averaging time of $\sim$200 s. After removing the long-term temperature and control electronics effects from the raw data, a linear drift of 268 $\mu$Gal/day is observed in the data, which is among one of the best reported for a MEMS device. These results demonstrate that this MEMS gravimeter is capable of conducting long-therm time-lapse gravimetry, a functionality essential for applications such as volcanology.

\section{Introduction}
Gravimetry has been used extensively over the past century in several fields. Due to the high cost of commercial instruments (around £70k for a portable device) the oil and gas industry has been the most prolific user of gravimetry, where it is often used as a pre-drilling survey method to investigate subterranean geological features \cite{Barnes2012,Rim2015}. Gravimetry has also been used, however, for sinkhole analysis \cite{Kaufmann2014}, finding tunnels and cavities for the defence sector \cite{Butler1984,Romaides2001}, CO$_2$ sequestration \cite{Gasperikova2008}, geothermal reservoir monitoring \cite{Nishijima2016}, archaeology \cite{Panisova2009}, hydrology \cite{Jin2013,Fores2017}, and volcanology \cite{Fernandez2017,Carbone2017,Aparicio2014,Battaglia2008,Rymer2000a}.

The last item in this list is perhaps the field in which gravimetry could offer the most societal benefit, but for which its use has been severely limited by the high capital cost of the equipment. Gravimetry is the only means by which mass variations within volcanoes can be measured. Gravimetry therefore offers a great advantage to hazard forecasting because mass changes often precede eruptive processes \cite{Calahorrano-DiPatre2019,Carbone2017,Poland2016}. Freire et al. \cite{Freire2019} state that `more than 8\% of the world’s 2015 population lived within 100 km of a volcano with at least one significant eruption'. The significance to humanity of improving eruption forecasting cannot therefore be overstated.  Gravimetry has been used to a limited extent in this setting, but spatial resolution has been limited by the number of instruments that can be affordably bought. Whilst single sensors can only provide a point measurement, if the price of gravimeters could be significantly reduced then multi-pixel gravity imaging would become possible. An analogy could be drawn between this and the example of a digital camera; with a single pixel you cannot capture a picture, you only have a light sensor, but this soon changes as you gain pixels. In 2019 Carbone et al. \cite{Carbone2019} described a measurement using only three sensors on Mt. Etna as an `network'; such is the sparsity of gravimeters on what is one of the most studied volcanoes in the world. With this application in mind, the authors have joined a consortium (NEWTON-g) to develop a gravimeter network on Mt. Etna comprising tens of MEMS gravimeter `pixels'\cite{Carbone2020}, and a MuQuans Absolute Quantum Gravimeter \cite{Menoret2018}.

Due to the inverse square law of gravity, a single measurement does not have a single unique solution; it is impossible to tell whether a signal has a source that is big and far away, or small and close. Inversion techniques therefore need to be applied to understand the data. Arrays such as the planned network on Mt. Etna present a great opportunity to reduce the problem of inversion, because by using two or more sensors one benefits from triangulation in reducing the number of possible solutions. The more sensors one introduces, the more information that it is possible to glean from the data. In essence, the data from an array of sensors is greater than the sum of its parts; and multi-pixel gravity imaging has the potential to change the way that mass changes within volcanoes are monitored.

\section{Design of the MEMS Gravimeter}

The MEMS gravimeter has two distinct components -- a monolithically etched silicon sensor with a high acceleration sensitivity; and an integrated capacitive readout scheme to measure the sensor's output. Here, the design aspects of the core MEMS sensor will be discsused briefly, followed by an explanation of the implemented readout scheme.

\begin{figure}
\centering
\includegraphics[width=1\textwidth]{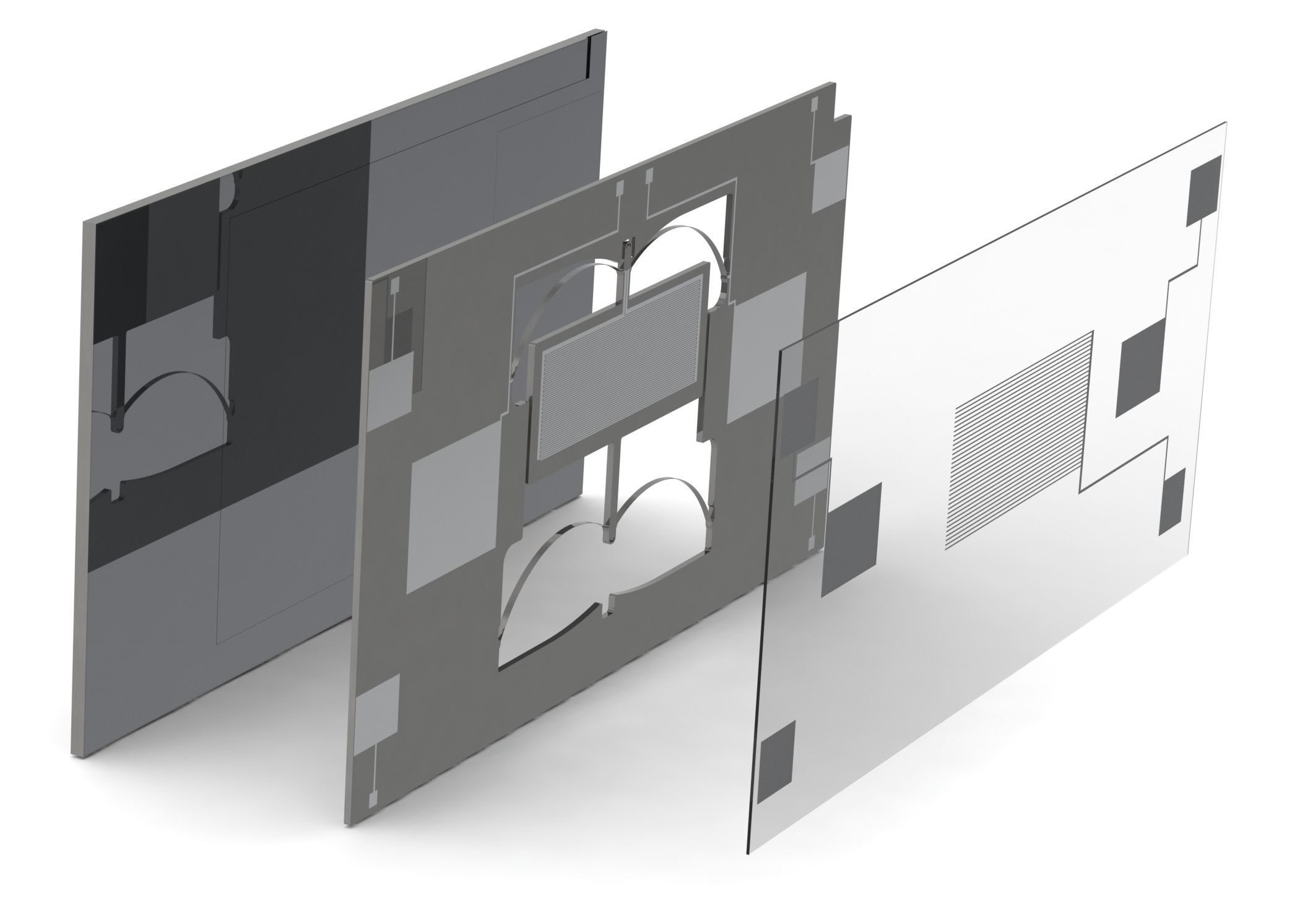}
\caption{\label{fig:Schematic}: The MEMS Device. An exploded view of the MEMS gravimeter assembly. The middle MEMS layer is mounted upon a base plate. The silicon base plate simply serves as a mechanical support for the MEMS layer. The MEMS layer has metal capacitive electrodes patterned on its surface. Corresponding pickup electrodes are patterned on an upper glass layer (made from SD2 silica), which is used to read out the signal as the MEMS proof-mass moves.}
\end{figure}

The design of the sensor is a based upon the device first presented by the group in 2016 \cite{Middlemiss2016,Middlemiss2016a,Campsie2016}. Significant changes have been made since this time, however, in order to miniaturise the sensor and improve both its sensitivity and stability. The sensor is a relative gravimeter: it is able to measure changes in gravity by monitoring the relative displacement of a proof-mass on a spring. For an oscillating system such as this, the input acceleration, $a$, and the corresponding proof-mass displacement, $z$, are directly proportional to each other through the relationship $a = 4\pi^2f^2 z$, where $f$ is the fundamental resonant frequency, and the bandwidth of the sensor. Here, the mass and the supporting springs are etched monolithicically from a single piece of silicon using standard photolithography and etching techniques \cite{Laermer2003} (see figure \ref{fig:Schematic}). The four symmetric springs utilise a geometric anti-spring design \cite{Bertolini1999,Cella2005,Acernese2015a,Abbott2016,ibrahim2008}. The details of how this design can be tuned for the purposes of gravimeter fabrication is detailed in the paper by Middlemiss et. al. \cite{Middlemiss2021}. This design enables a decreased spring stiffness in the vertical direction whilst limiting the motion in the horizontal and out-of-plane axes. Decreasing the stiffness -- and hence the oscillation frequency -- in the operation axis is important because it means that a proof-mass will move more for a given change in gravity. This is particularly useful in cases where the sensitivity of the system is limited by the displacement readout. Softening the spring, however, comes at a cost because the robustness of the sensor decreases as the resonant frequency drops. With the intention of developing a much more robust sensor, the spring design was optimised to obtain a working resonant frequency of $7.35$ Hz (compared to $2.2$ Hz in earlier publications \cite{Middlemiss2016}). Given that there was also a desire to improve on the sensitivity reported in the earlier work, it was necessary to improve the displacement sensitivity measurement of the proof-mass.

The measurement of the MEMS displacement is, of course, another fundamental function of the device. In Middlemiss et. al. \cite{Middlemiss2016}, an optical shadow sensor was used to measure the displacement of the mass \cite{Bramsiepe2018a}. An LED was used to cast light on a photodiode, and the MEMS was placed in the light beam. As the mass moved, the shadow cast on the photodiode altered the photocurrent, which was used as a proxy for displacement. This methodology was limiting for several reasons. A displacement sensitivity of less that 1 nm was difficult to achieve, ultimately limiting the acceleration sensitivity of the overall sensor. The components needed to be mounted on a bespoke machined block of fused silica to reduce temperature sensitivity. The disadvantage of this block was twofold: it was expensive to machine, and it was far too big to fit inside a standard MEMS package. Whilst this device configuration was used to conduct field tests \cite{Middlemiss2017,Middlemiss2018,Prasad2018}, further minaturisation was required.

To facilitate the device miniaturisation, a capacitive displacement method  was adopted to readout the motion of the proof-mass. A set of metal comb electrodes were patterned on the top surface of the proof-mass while a complementary set of electrodes were patterned on a fixed glass layer that was separated from the proof-mass layer by a gap of 40 $\mu$m using SU8 spacers. The combs on the two layers formed overlapping capacitors. While the proof-mass comb was electrically driven by a 40 kHz differential pair of sine waves, the signal was picked up by the set of fixed combs on the glass layer. Any displacement of the proof-mass modulates the overlapping capacitance and, thus, the output current flowing through the capacitors. Therefore, by monitoring the changing overlapping capacitance, the position of the mass could be measured. The current obtained at the output of the overlapping capacitors was fed to an operational amplifier that was configured in the conventional transimpedance amplification (TIA) topology. The transimpedance topology, despite its simplicity, is very effective in reading current from high impedance sources because its signal gain is insensitive to capacitive parasitics at the pick-up stage\footnote{The noise gain of the TIA stage, however, is a function of the input parasitics and needs more attention during circuit design when the sensitivity of the system is limited by the electronics noise floor.}. The voltage signal obtained after the amplification stage was then fed to a bench-top lock-in amplifier. The signal was demodulated and low-pass filtered, and the signal amplitude that carried the displacement information was acquired and logged onto a PC through a Labview routine at a sampling frequency of roughly 0.2 Hz. Using the modulation-demodulation lock-in approach to extract the signal amplitude helps in lowering the noise floor as it only allows the noise at the carrier frequency to pass through while rejecting the noise at all other frequencies.

The capacitive displacement measurement mechanism is advantageous both because it offers a significant reduction in size compared to the shadow sensor (both the sensor and the readout were implemented on the same chip stack), but also because it had a displacement sensitivity that was significantly better than the optical shadow sensor. With a capacity to resolve 50-100 pm of proof-mass displacement, the capacitive displacement readout represented a sensitivity improvement over the shadow sensor by a factor of 10-20. This readout methodology could be improved even further because the displacement sensitivity of the capacitive readout ($\dfrac{\delta C}{\delta z}$, where $C$ is the overlapping capacitance) is a function of the comb geometry as well as that of the gap between the two overlapping combs\cite{heerens1982multi}. By increasing the comb finger density and bringing the proof-mass and the glass pick-up layer closer, the displacement sensitivity can be enhanced. In the case of our sensor, however, we opted for design parameters (comb density, and gap) that provided a significant safety margin on the fabrication and assembly tolerances. This, of course, can be optimised in the future.

To ensure the long-term stability and the noise performance of the MEMS gravimeter, it was necessary to control the temperature of the device. As discussed in Middlemiss et. al. \cite{Middlemiss2016}, temperature fluctuations alter the Young's Modulus of the silicon flexures, which in turn alters the spring constant of the MEMS device; which can cause spurious readings if not accounted for. Multiple resistance temperature detectors (RTDs) were therefore used to monitor the temperature of the instrument at multiple locations.  The gravimeter was fixed on to a peltier element to control the temperature within a milli Kelvin of the nominal temperature set-point. More detail on the measurement set-up are provided in the Methods section of the manuscript.

\section{Results}

\subsection{Earth Tide Measurement}
One means of assessing the performance of a gravimeter is to measure the Earth tides. The Earth tides are caused by tidal forces in the Sun-Earth-Moon system \cite{Farrell1973}. The relative phase of the Sun and Moon cause the Earth to deform elastically. The effect of this deformation is a diurnal and semi-diurnal periodic variation in the distance between the crust and the centre of mass of our planet, and thus a corresponding variation in the gravitational acceleration measured at any given location on Earth. This periodic signal has an amplitude that varies between around 100 $\mu$Gal and 300 $\mu$Gal, depending on the monthly phase of the Moon. Maximum amplitude signals are referred to as `spring tides' and minimum amplitude signals are referred to as `neap tides'. A measurement of the Earth tides is an ideal means of demonstrating gravimeter performance because such a measurement demonstrates both stability and sensitivity of the instrument in question. Furthermore, it is a globally recognised signal in the field of gravimetry, and it can be readily calculated using software such as T-Soft \cite{van2005uncertainty}. T-Soft can also be used to account for the effect of ocean loading: a local variation in the Earth tide signal caused by compression and subsequent rebound of land due to the tidal motion of large bodies of water. Ocean loading effects are more complex to calculate because they require knowledge of local topology and fluid dynamics, and can alter both the amplitude and phase of Earth tide signals by around 5$\%$.

Previous measurements have been made of the Earth tides with MEMS gravimeters. In 2016 Middlemiss et. al. published a measurement lasting six days \cite{Middlemiss2016}, with a correlation coefficient of 0.86 between the experimental data and a theoretical Earth tide calculated with T-soft. This correlation coefficient was calculated using the Pearson product-moment correlation coefficient. More recently, Tang et. al. \cite{Tang2019} produced a device based on that of Middlemiss et al. This device was used to make a 5.5 day measurement of the Earth tides, with a correlation coefficient of 0.91 when compared to a co-located commercial superconducting gravimeter.

\begin{figure}
\centering
\includegraphics[width=1\textwidth]{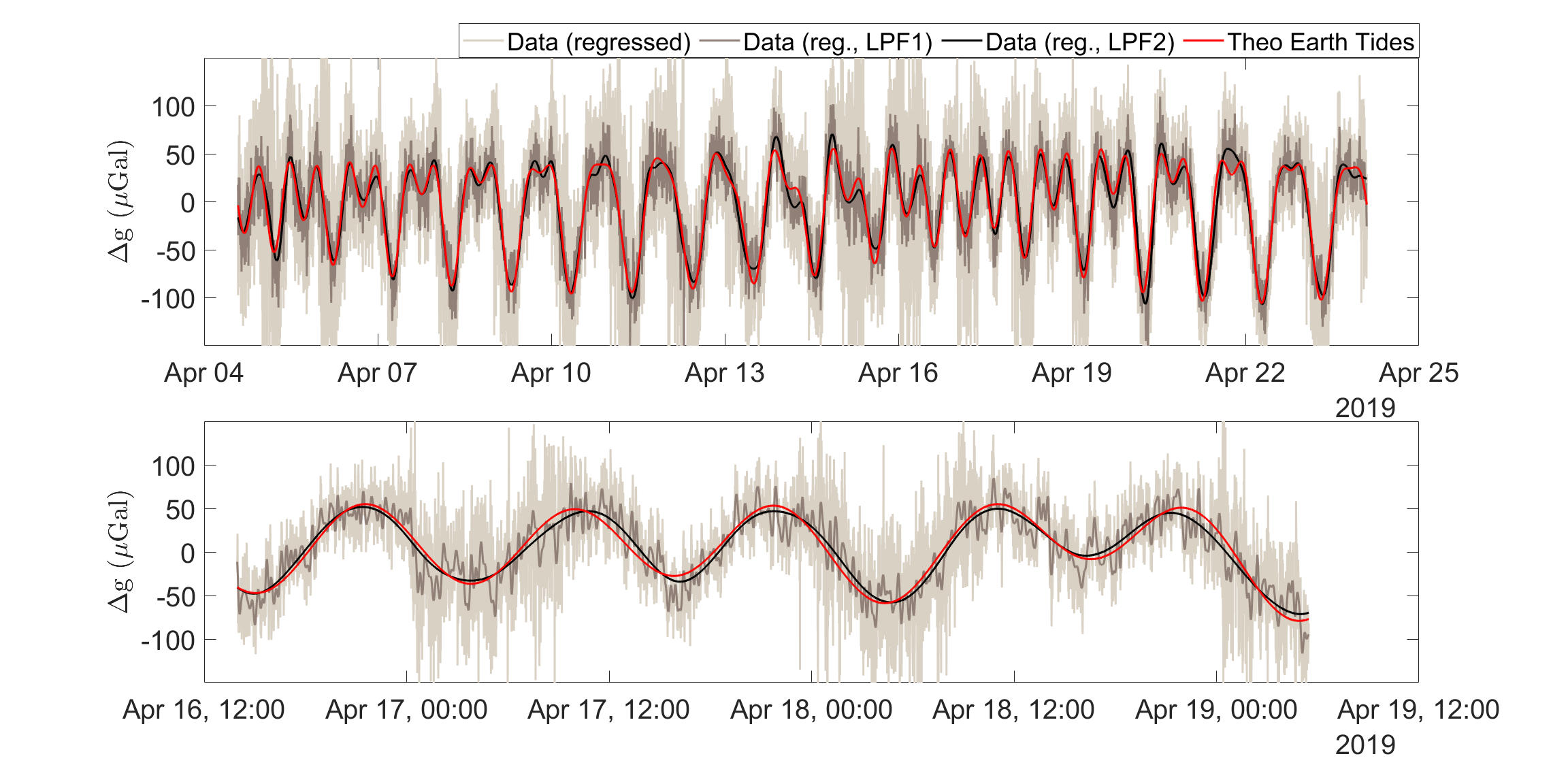}
\bf{\caption{\label{fig:Tides}{{: Two time series plots of the Earth tide measurement, conducted at Glasgow University in April 2019. In both graphs, the red series is a theoretical signal calculated using T-Soft, the gray-scale series are the experimental data recorded using the MEMS gravimeter. From lightest grey to black, these series represent increasing filtering of the data using a low-pass filter (LPF). The black series was filtered with a 20 $\mu$Hz bandwidth LPF. All of the theoretical data is presented after a regression analysis was used to remove drifts caused by temperature variations within the system. The upper graph shows the full data set, the lower graph shows a zoomed in selection of this data.}}}}
\end{figure}

Using the MEMS gravimeter described above, a measurement of the Earth Tides was made. This data, starting on the 4th of April 2019, can be seen in figure \ref{fig:Tides}. The measurements were taken at a sample rate of $0.18$ Hz. The red series is a theoretical signal calculated using T-Soft for the measurement location (Glasgow, U.K). This theoretical series includes a correction for ocean loading using the GOT00.2 model \cite{ray1999global}. The light grey series is the unfiltered, post-regression acceleration data from the MEMS device (see Methods section for regression analysis details). The dark grey series shows the acceleration data after the application of a low-pass filter (LPF) with a bandwidth of 1 $m$Hz. The solid black line shows the same data after a final stage of filtering, this time using a 20 $\mu$Hz bandwidth LPF. There is a correlation coefficient, $R$, of 0.979 between the theoretical data and the 2nd-stage filtered data (calculated using the Pearson product-moment correlation coefficient).

\subsection{Spectral Analysis}

\begin{figure}
\includegraphics[width=1\textwidth]{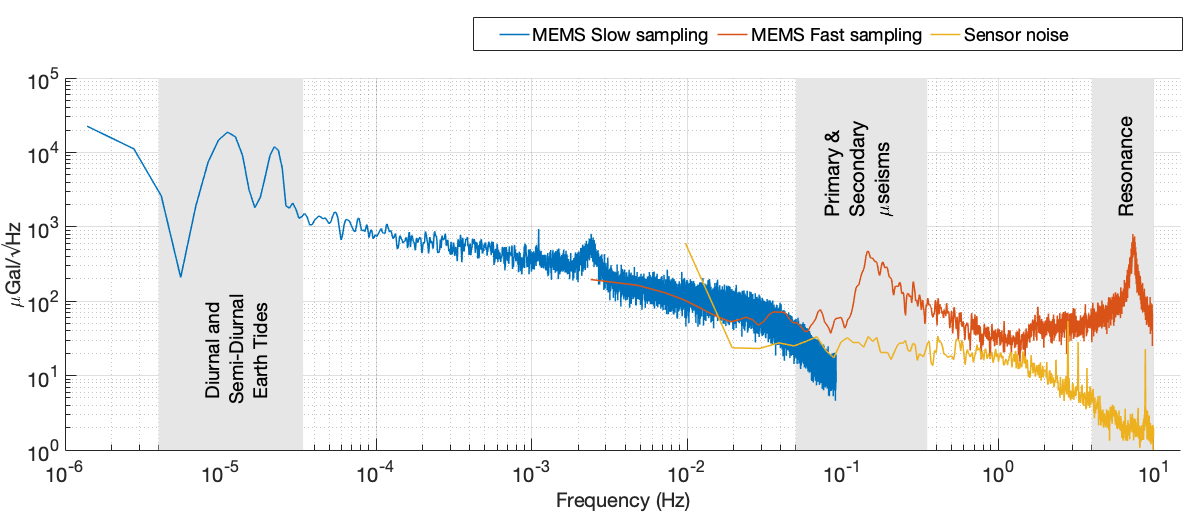}
\centering
\caption{\label{fig:ASD}: An amplitude spectral density plot of the post-regression instrument data (the blue and the orange series), and the instrument noise data (the yellow series). The blue series represents data that were recorded at a sampling rate of 0.18 Hz. The orange and the yellow series were recorded at a sampling rate of 20 Hz. Three distinct frequency bands of interest can be observed in the data - these are highlighted using the grey bands. At 10$^{-5}$ Hz, the diurnal and semi-diurnal components of the Earth tides are visible; the primary and secondary microseisms can be observed at around 0.2 Hz; and the resonant peak of the MEMS device itself is located at 7.35 Hz. The sensor noise has a flat spectral response for the measured bandwidth. \textit{Note:} The downward roll-off at high frequencies for the blue and the yellow series is due to post-demodulation low-pass filtering.}
\end{figure}

The spectral analysis of the gravimeter data provides further insights into the presence of periodic signals of varying frequencies in the acceleration data. The blue series of figure \ref{fig:ASD} is a presentation of the drift-corrected, regressed data (the light grey series of figure \ref{fig:Tides}) in the form of an amplitude spectral density graph. To complement the slow-sampled gravimeter data, a second series of higher-frequency gravimeter data (sampled at a rate of $20$ Hz) is also included in the plot\footnote{The gravimeter data was not sampled at a higher sampling rate for the entire bandwidth to avoid unnecessary accumulation of data.}. This can be observed in the orange series in figure \ref{fig:Tides}. The yellow series in figure \ref{fig:Tides} is the electronic noise floor of the sensor.

The blue series in figure \ref{fig:Tides} has a clearly identifiable double peak at $10^{-5}$ Hz. These two peaks represent the diurnal, and semi-diurnal periodicity of the Earth tides. This signal splitting occurs as the relative phase of the Sun-Earth-Moon system changes. When all are aligned in a line, the gravitational signal of the Sun and the Moon acts in phase on the Earth. When these three bodies form a right-angled triangle, however, the gravitational signal of the Sun and Moon acting on the Earth are out of phase. It is this same phase drift that causes the total amplitude to vary between spring and neap tides. Another demonstration of this Earth tide signal splitting has not before been presented in the literature (to the knowledge of the authors). Such data has not previously been presented because its observation requires the a clean time series of well over a week. In addition to the double-peak, there is a small spurious peak occurring around 25 $m$Hz. This signal is not geophysical in nature, and has been identified as being caused by the temperature control electronics. The roll-off seen in the slow-sampled data at higher frequencies is due to the low-pass filter operation performed by the lock-in amplifier.

In the orange series, further signals can be observed. The furthest right of these is the resonance peak of the MEMS mass-on-spring system at $7.35$ Hz. This represents the maximum frequency at which the device can record meaningful data. The sensitivity to signals of higher frequencies than this resonance will rapidly decrease due to the nature of a simple harmonic oscillator. The second set of signals that can be identified in the orange series is geophysical in nature: the primary and secondary mircoseisms. The primary microseism occurs at frequencies between 0.04 and 0.15 Hz \cite{essen2003generation}, and is caused by interactions between ocean waves and the sea bed \cite{Lepore2018}. The secondary microseism has a larger amplitude, and occurs between 0.08 and 0.3 Hz \cite{essen2003generation}. This signal is also related to ocean activity, but is caused by interactions in shallow water between wind-driven ocean waves, and waves reflected back from the coastline. Both the primary microseisms vary in amplitude and phase, depending on ocean conditions.

As the higher-frequency band of the sensor output is dominated by the ground motion, it is difficult to estimate the true noise floor of the sensor. One way to get rid of this issue is to use a reference seismometer to effectively subtract the sesimic signals in the high-frequency band. However, we did not have access to such an instrument. We instead estimated the sensitivity by measuring the true electronic noise floor of the full sensor set-up. This was achieved by grounding the modulation sine wave drive at the input of the MEMS capacitors. The measured sensor noise data is represented by the yellow series in the figure, and shows a flat spectral response for the measured frequency bandwidth. The roll-off observed for higher frequencies, again is due to the low-pass filtering. The sensitivity of the gravimeter is estimated to be 18 $\mu$Gal/$\sqrt{Hz}$ at 1 Hz, but at this frequency the sensor is limited by seismic noise. As discussed in the next section, the ultimate noise floor of the sensor is 0.91 $\mu$Gal, which can be achieved over an integration time of 250 s.

It should also be noted that the device is also a sensitive seismometer. One way in which his can be observed is via the presence of anthropogenic noise in \ref{fig:Tides}. It is quite evident that the amplitude of the high-frequency components increases and then decreases with the changing footfall within the building. Since it is difficult to capture this time-localised information using the standard fourier analysis, we have used wavelet analysis to isolate the periodic variation in the anthropoegenic noise. The details of the analysis are included in the Appendix. 

\subsection{Stability}
\subsubsection{Allan Deviation Analysis}
\begin{figure}
\includegraphics[width=1\textwidth]{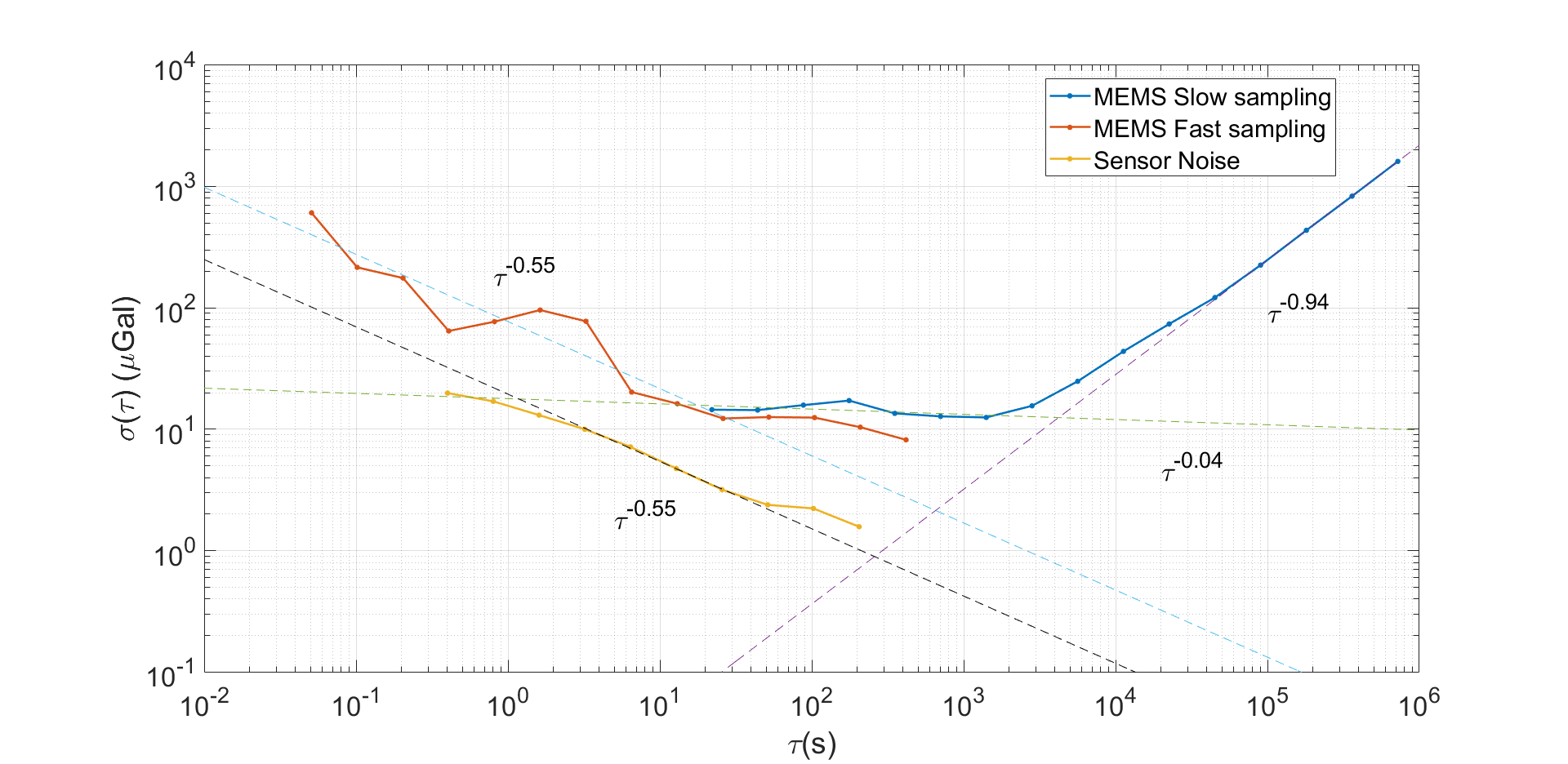}
\centering
\caption{\label{fig:Allan}: The Allan Deviation (AD) plots for the raw slow sampled MEMS data (the blue series), the raw fast sampled MEMS data (the orange series), and the sensor noise data (the yellow series). Dashed series are a result of fitting straight lines across different averaging periods for each data set. The slope of each fit is represented in the form of power-laws. \textit{Note:} The small integration time ($\tau$) data is not shown for the blue ($\tau < 2^{2}t_{\text{slow}}$, where, $t_{\text{slow}} = 5.5$ sec.) and the yellow series ($\tau < 2^{3}t_{\text{fast}}$, where, $t_{\text{fast}} = 0.05$ sec.) as the AD values for these times are affected by the low-pass filtering stage.}
\end{figure}

To understand the stability and the noise characteristics of the gravimeter, the Allan-Deviation (AD) values ($\sigma$) were computed from the raw, uncompensated gravimeter data.\footnote{The Allan deviation (or, variance) approach, which was originally developed to quantify the stability of clocks in the time-domain \cite{allan1975measurement}, is now routinely used to extract the stability metrics of IMUs (Inertial Measurement units)\cite{el2007analysis}.} The $\sigma$ values for the slow- and the fast-sampled data, and the sensor noise are plotted in Figure \ref{fig:Allan}. We have used the fast-sampling data (the orange series in figure \ref{fig:ASD}) for evaluating the stability of the gravimeter for shorter sampling/averaging periods ($\tau$), and the slow-sampling data (the blue series in figure \ref{fig:ASD}) for assessing the long-term stability. In the case of the slow-sampled data, the first few $\tau$ values ($\tau =$ 5.5 s, 11 s) have been excluded because the low-pass filter was shaping the frequency response. As the fast-sampled data was logged for a sufficiently long duration, there were enough samples in the fast-sampled data to compute the $\sigma$ values for the missing $\tau$ values in the slow-sampled series. For the same reason, the first few $\tau$ values ($\tau =$ 0.05 s, 0.1 s, 0.2 s) have been excluded from the sensor noise data as well.

Interpreting the noise characteristics from the AD plots is straightforward. For example, in the case of smaller averaging periods (fast sampling MEMS data, $\tau<$ 30 s), the $\sigma$ values continue to decrease, underscoring the advantage of averaging to achieve to the best sensitivity of the gravimeter. Fitting a straight line through the data (dashed red line) produces a slope of $-0.55$, indicating the dominance of white noise\cite{el2007analysis} for these integration times. The noise in this band is a combination of uncorrelated random walk (white noise), and correlated noise arising from the device resonance, high-frequency anthropogenic noise and the microseismic ground motion. For higher averaging periods (fast and slow sampling data, 30 s $<\tau<$ 2000 s), the band is dominated with flicker noise evident from the slope ($-0.04$) of the fitted straight line (dashed blue line). As the $\sigma$ values remain mostly unchanged in this band, there is no advantage of averaging the data further. The deviation values in this band are also the smallest considering all the possible averaging periods, and determine the \textit{bias instability} of the MEMS device. The smallest $\sigma$ value, and, hence, the bias instability of 8.18 $\mu$Gal is obtained at $\tau\sim$ 417 s when considering the fast-sampled data. Taking the slow-sampled series into account as well, the bias instability is never worse than 17.22 $\mu$Gal (the $\sigma$ value at $\tau\sim$ 176 s). An average of all the $\sigma$ values in the band, for both the slow- and fast-sampled data, gives a \textit{mean} bias instability figure of 13.22 ($\pm$2.73) $\mu$Gal between an averaging period range of 30 s to 2000 s. The higher averaging periods ($\tau>2000s$) are dominated by sensor drift, mostly caused by the temperature related effects and the control instrumentation. Fitting a straight line (dashed violet line) through the last few $\sigma$ values reveal a slope of close to 1 ($0.94$) that is consistent with the drift rate ramp behaviour\cite{el2007analysis}. The worst drift rate of 216 $\mu$Gals/day for the device is captured at $\tau\sim$ 1.04 days where $\sigma$ sits at 224.67 $\mu$Gal.

The AD values obtained using the sensor noise represent the best-case scenario and can potentially be used to predict the noise characteristics of the gravimeter in the absence of the usually ubiquitous ground motion. In the case of sensor noise, the $\sigma$ values continue to decrease with a slope of -0.55, obtained from fitting a straight line (the black dashed line in the figure), and reaches a minimum value of 1.6 $\mu$Gal for a $\tau$ of $\sim$205 s. By extrapolating both the drift ramp from the slow sampled data (dashed blue line) and the sensor noise (dashed black line), the intersection point of the two noise regimes can be obtained. In our case, that intersection point occurs at $\tau\sim$ 250 s, with an AD value of 0.91 $\mu$Gal, that is lower than the bias instability measured just using the sensor noise data. Hence, in the absence of any ground motion, we can conclude that the ultimate bias instability noise floor is 0.91 $\mu$Gal.

\subsubsection{Drift characteristics}
\begin{figure}
\centering
\includegraphics[width=1\textwidth]{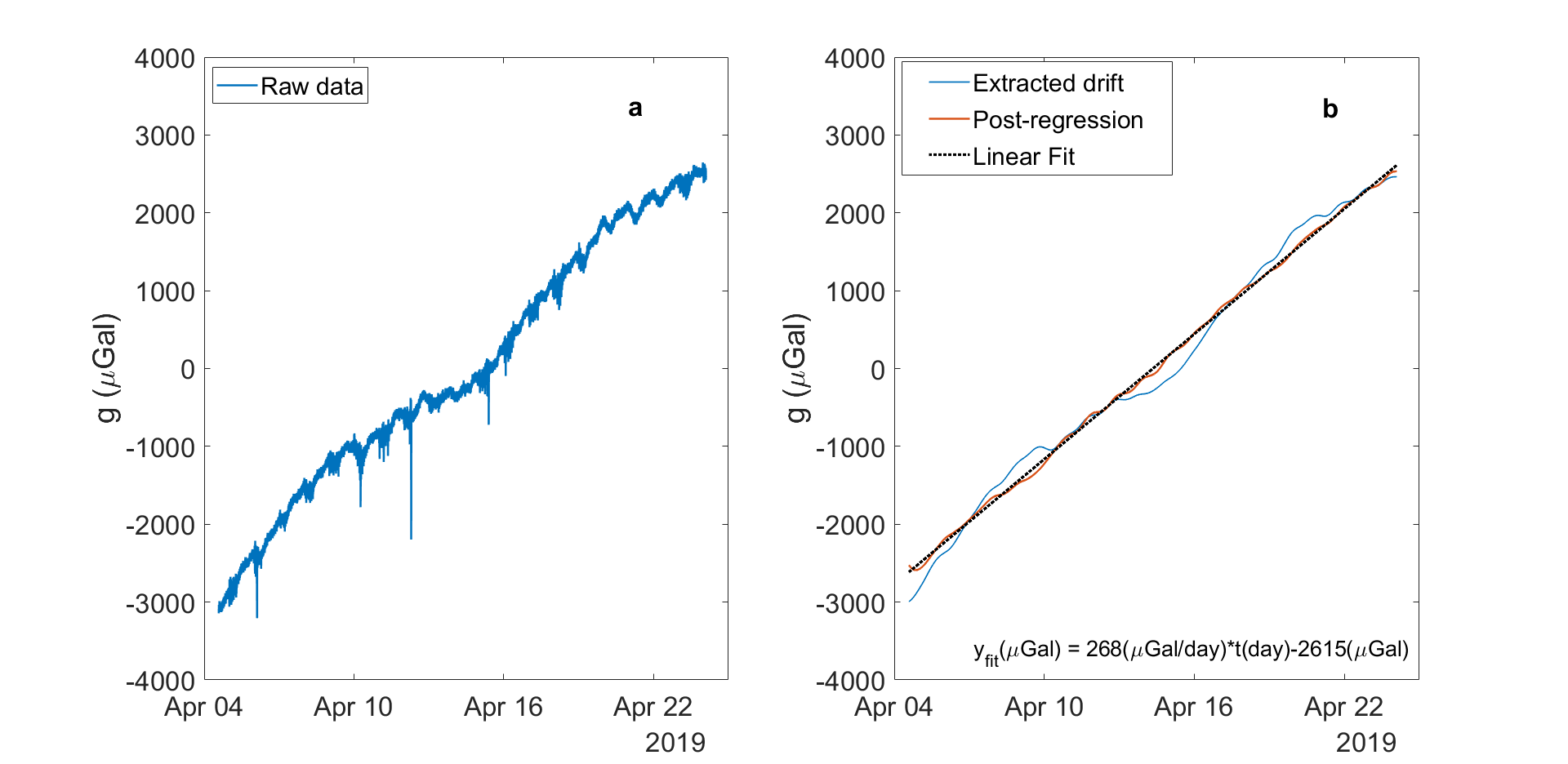}
\caption{\label{fig:raw}: The raw data (a) and the drift analysis (b) plots for the gravimeter. The raw data is first low-pass filtered to retain only very long-term features ($<$2 $\mu$Hz) in the raw data (blue series in (b)). The extracted drift is then processed to regress out the impact of the temperature and the control instrumentation. The post-regression data (red series) is then fitted against a straight line (dashed black series) to obtain the drift rate of the gravimeter.}
\end{figure}

As it is difficult to visualise the hidden features in the the drift of the device by simply computing the AD values, we used regression analysis and polynomial curve fitting on the raw, uncompensated data to study the drift behaviour. The raw sensor data is plotted in Figure \ref{fig:raw}(a) and clearly shows a non-linear behaviour (the rounded `M' shape) on top of a linear trend. To remove the tides from affecting the analysis, we applied low-pass filtering, with a 2 $\mu$Hz cut-off, to the data. Setting this bandwidth was a compromise between avoiding filtering the non-linear aspects of the drift while attenuating the tide signal enough from the series. The effect of filtering can be seen in Figure \ref{fig:raw}(b), where the low-pass filtered data is represented by the blue series. In the next step, we regressed the filtered data against the instrument's temperature and PID control output channels. Correcting for the long term temperature effects resulted in the removal of the non-linear trend leaving an almost linear drift (the red series) in the data. We next fitted a straight line to the corrected data and obtained a linear drift of $268\mu$Gal/day with an $r^2$ coefficient of 0.99.\footnote{The drift rate obtained using this analysis is slightly higher compared to the one obtained using the AD approach, and could be attributed to less than ideal signal processing (regression, fitting) of the data.} It is believed that the origin of this trend is environmental (for example, temperature gradient effects that are not captured through the temperature sensors), and not geophysical.

\section{Conclusions}
In this paper, a 19 day measurement of the Earth tides has been presented. The correlation coefficient between this measurement and a theoretical signal is 0.96, demonstrating a remarkable stability for a MEMS sensor. This sensor utilises an in-plane capacitive displacement readout, which has reduced the size of the device compared to previous work by the authors. This displacement sensor has also been used to increase the displacement sensitivity to 40-50pm, leading to a bias instability value of $8.18 \mu$Gal. A network of these sensors is currently being constructed for deployment on Mount Etna as part of the NEWTON-g collaboration \cite{Carbone2020}, to create the first multi-pixel gravity imager.

\bibliographystyle{naturemag2}

\clearpage
\section*{Acknowledgements}

The authors would like to thank Kelvin Nanotechnology who fabricated the final MEMS sensors discussed in this paper. We would also like to thank the staff and other users of the James Watt Nanofabrication Centre for help and support in undertaking the MEMS fabrication development.

This work was funded by the Royal Society Paul Instrument Fund, STFC grant number ST/M000427/1, and the UK National Quantum Technology Hub in Quantum Enhanced Imaging (EP/M01326X/1), the EU H2020 project `NEWTON-g' (H2020-FETOPEN-1-2016-2017) and the Royal Academy of Engineering (Project RF/201819/18/83)

\section*{Author contributions}

\begin{itemize}

\item{A. Prassad and R. P. Middlemiss are credited as joint first authors of this manuscript.}

\item{A. Prassad led the data analysis to correct for the drift in the data, and co-wrote the manuscript.}

\item{R. P. Middlemiss led the fabrication of the MEMS device, conducted data analysis, contributed to the initial conceptual design of the gravimeter, and co-wrote the manuscript.}

\item{K. Anastasiou contributed to the thermal control of the MEMS device.}

\item{S. Bramsiepe contributed to the electronic readout of the device.}

\item{A. Noack contributed to the design of the capacitive readout of the MEMS device.}

\item{D. J. Paul contributed to the initial conceptual design of the gravimeter, and contributed to the manuscript preparation.}

\item{K. Toland contributed to the analysis of anthropogenic noise observed in the data.}

\item{P. R. Utting conducted the cross correlation analysis of the tidal time-series.}

\item{G. D. Hammond had oversight of the experimental design, led the initial conceptual design of the gravimeter, conducted data analysis, and contributed to the manuscript preparation}.

\end{itemize}

\section*{Author information}
\begin{itemize}
\item{The research data relevant to this paper will be stored on the University of Glasgow's Enlighten Repository on publication}
\item{The authors have no competing financial interests.}
\item{Correspondence and requests for material should be addressed to abhinav.prasad@glasgow.ac.uk, richard.middlemiss@glasgow.ac.uk, or giles.hammond@glasgow.ac.uk}
\end{itemize}

\end{document}